\documentclass[sigconf,nonacm]{acmart}
\usepackage{longtable}
\usepackage{subcaption}
\AtBeginDocument{%
  }

\begin{document}

\title{Exploiting Feature-invariant Scaling Laws for Efficient A/B Experimentation in Recommendation Systems}

\author{Zhiliang Chen}
\affiliation{%
  \institution{Meta, National University of Singapore}
  \country{}
}
\email{chenzhiliang@u.nus.edu}

\author{Sammy Nammari}
\affiliation{%
  \institution{Meta}
  \country{}
}
\email{nammari@meta.com}

\author{Jiahao Su}
\affiliation{%
  \institution{Meta}
  \country{}
}
\email{jiahaosu@meta.com}

\author{Xiaoqiang Luo}
\affiliation{%
  \institution{Meta}
  \country{}
}
\email{xql@meta.com}

\renewcommand{\shortauthors}{Chen et al.}

\begin{abstract}
Researchers and engineers improve recommendation systems by performing A/B experiments with new model architectures, optimization methods, and data features. However, running full-scale experiments (e.g., full-model training with online data) for every feature change is prohibitively computationally intensive, while the result from a single small-scale experiment is often misleading. To address this challenge, we show that modern recommendation models exhibit \textit{feature-invariant scaling}: the change of a model's performance follows a predictable scaling-law even when different feature changes are applied to the baseline model. We introduce an efficient method that exploits scaling-laws and uses statistical inference to predict (with confidence intervals) the full-scale effect of an A/B experiment in recommendation systems. This enables practitioners to efficiently filter out neutral or regressive feature changes with only small-scale experimentation. We validate our method on three distinct recommendation model families---Wukong, HSTU, and LLM-based recommendation models---across more than 40 architectural, optimization, and data changes. Our method accurately distinguishes beneficial feature changes from neutral and regressive ones while using only a fraction of the compute required for full-scale evaluation. It also identifies several changes that improve HSTU's state-of-the-art performance on MovieLens-20M. Our results demonstrate that scaling laws can be practically exploited to make compute-efficient decisions for improving recommendation systems.
\end{abstract}




\begin{teaserfigure}
 \includegraphics[width=1.0\textwidth]{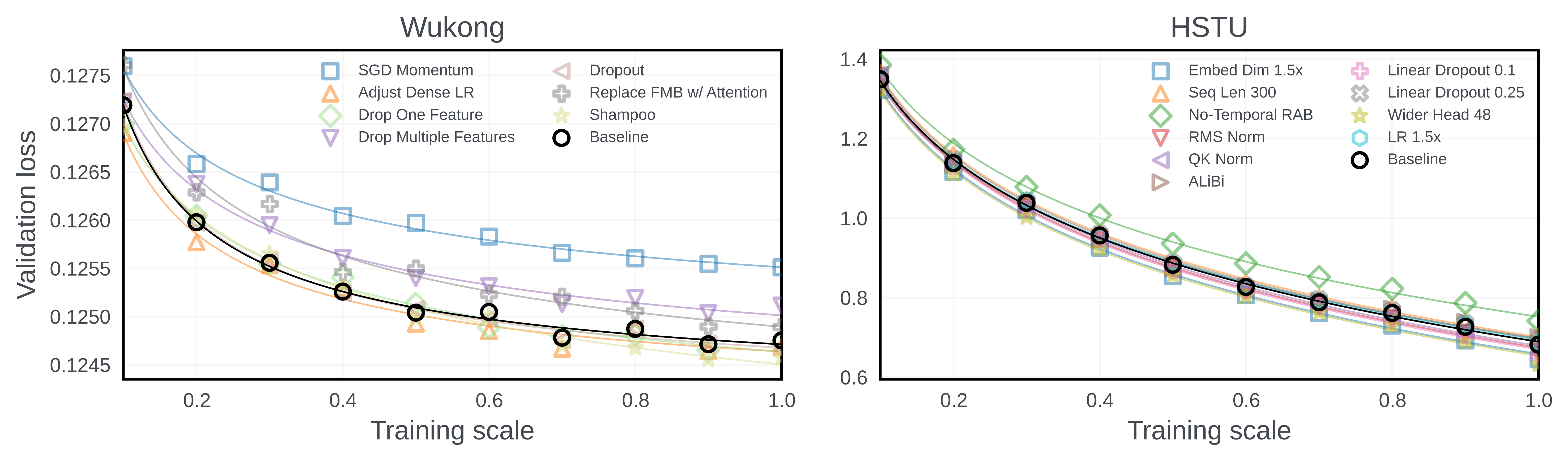}
\caption{Feature-invariant scaling laws for Wukong \cite{zhang2024wukongscalinglawlargescale} and HSTU \cite{zhai2024actionsspeaklouderwordsHSTU} over a wide variety of model, data, and optimizer features added. Regardless whether a feature improves or regresses the model, the model validation loss changes predictably with training scale.}
\label{fig:feature-invariant-scaling-wukong-hstu-front}
\end{teaserfigure}

\maketitle

\section{Introduction}
Researchers and engineers often develop \textit{new feature changes} (e.g., architecture changes, new optimization techniques, new data features) to improve the prediction accuracy of a recommendation model, which serves as a major driver for many large-scale online platforms. An important part of this process is to perform A/B experimentation to ensure that a new feature actually improves the model performance. This often requires a large amount of compute resources and time. For example, to know whether an architecture change improves a recommendation model, we need to retrain the full model on the full training data and compare it against the baseline model performance.

A popular approach is to run small-scale experiments (i.e., a small-sized model trained on a subset of the data) to assess the effectiveness of a new feature change before committing to a full-scale run~\cite{surrogate_1,surrogate_2,surrogate_3}. While these experiments offer some useful early signals, a single small-scale experiment is often an unreliable predictor of full-scale performance. An illustrative example is shown in Fig.~\ref{fig:motivating-example}, where new item information given to an LLM-based recommendation model degrades performance at a small scale but improves performance at full scale (see Sec.~\ref{sec:motivating-case-study} for in-depth analysis). Hence, relying on small-scale results alone can lead us to reject a beneficial feature change (or vice versa).

In this paper, we propose an efficient method that exploits \textit{scaling laws} to improve the reliability of A/B experimentation in recommendation systems. Our approach exploits an important insight that we introduce in this paper: the change in performance of modern recommendation models with increasing data and model scale is governed by predictable scaling laws. More importantly, the \textit{functional form} of these scaling laws remains consistent across a wide range of feature changes. As such, we can fit a scaling curve to a handful of small-scale experiments to predict the outcome of a full-scale A/B experiment. Specifically, our contributions are:


\begin{enumerate}
    \item We show that modern recommendation models enjoy \textit{feature-invariant scaling laws}, where their performance varies predictably with increasing data and model scales. In addition, the functional form of these scaling curves remains consistent across different new features (e.g., architecture changes, new optimization techniques, new data features).

    \item We introduce an efficient method that leverages a series of small-scale experiments and feature-invariant scaling laws to improve the reliability of small-scale experimentation. Our method uses statistical inference to infer the effectiveness of a new feature at the full-scale. This allows us to predict the effect of new feature changes without performing intensive full-scale experimentation. We also show that, in general, the cumulative resource use of these small-scale experiments is smaller than that of a full-scale experiment.
    \item We show that feature-invariant scaling laws hold true for different classes of recommendation models, including Wukong \cite{zhang2024wukongscalinglawlargescale}, HSTU \cite{zhai2024actionsspeaklouderwordsHSTU}, and GR2 (an LLM-based recommendation model) \cite{liang2026gr2generativereasoningreranker}. Then, we show that our method makes reliable predictions for a large range of new feature changes on these models (in determining if they improve or degrade performance). In this process, we also found several new features that improve the state-of-the-art performance of recommendation models on existing open-source datasets.
\end{enumerate}

\section{Related Work}

Extensive research has been devoted to developing new feature changes for recommendation models (and ML models in general). Examples include architectural changes \cite{zhai2024actionsspeaklouderwordsHSTU, naumov2019deeplearningrecommendationmodelDLRM, zhang2024wukongscalinglawlargescale}, interaction modeling techniques \cite{DCN}, optimization methods \cite{xie2021contrastivelearningsequentialrecommendation, PLE, gupta2018shampoopreconditionedstochastictensor}, and more. To determine if a new feature change is ``good'', a common practice in an industry setting is to add it to the baseline model and retrain the model from scratch with the entire set of (offline or online) training data. This is also colloquially known as full-scale A/B experimentation \cite{offline_ab, quin2023abtestingsystematicliterature}.

The major drawback of full-scale A/B experimentation is that it is computationally costly, especially when full-scale retraining is required. In addition, it must be repeated for every candidate feature change (which can be in the hundreds in an industrial setting). One naive approach is to make use of \textit{small-scale experiments}, which aim to build faithful surrogate small-scale models \cite{chen2026chickeneggdilemmacooptimizing, kadra2023scalinglawshyperparameteroptimization} that allow for fast iterations and reject ineffective feature changes if they do not show performance improvement at small scale \cite{surrogate_1, surrogate_2, surrogate_3}. However, a single small-scale experiment often produces misleading signals (see Sec.~\ref{sec:motivating-case-study}). As such, one could draw a wrong conclusion about the effectiveness of a new feature change, forgoing potential performance improvement (and revenue) in recommendation systems. Our work aims to circumvent such pitfalls associated with conventional small-scale experimentation.

\textit{Scaling laws} \cite{kaplan2020scalinglawsneurallanguage, hoffmann2022trainingcomputeoptimallargelanguage, zhang2024wukongscalinglawlargescale} describe how the performance of recommendation models or LLMs changes with increasing model and training data scale. Most existing literature merely describes the existence of these scaling laws. As far as we know, our paper is the first to show that a recommendation model with a new feature change applied \textit{still follows a scaling law with the same functional form} but different coefficients. In this paper, we refer to this phenomenon as \textit{feature-invariant scaling}. This property makes scaling laws useful for large-scale A/B experimentation. Because the functional form is known in advance, we can perform a few small-scale experiments and extrapolate the full-scale effect of \textit{every} candidate feature change, allowing us to filter out ineffective changes cheaply.

\section{A Motivating Case Study}\label{sec:motivating-case-study}
\begin{figure}[h]
  \centering

 \includegraphics[width=0.7\linewidth]{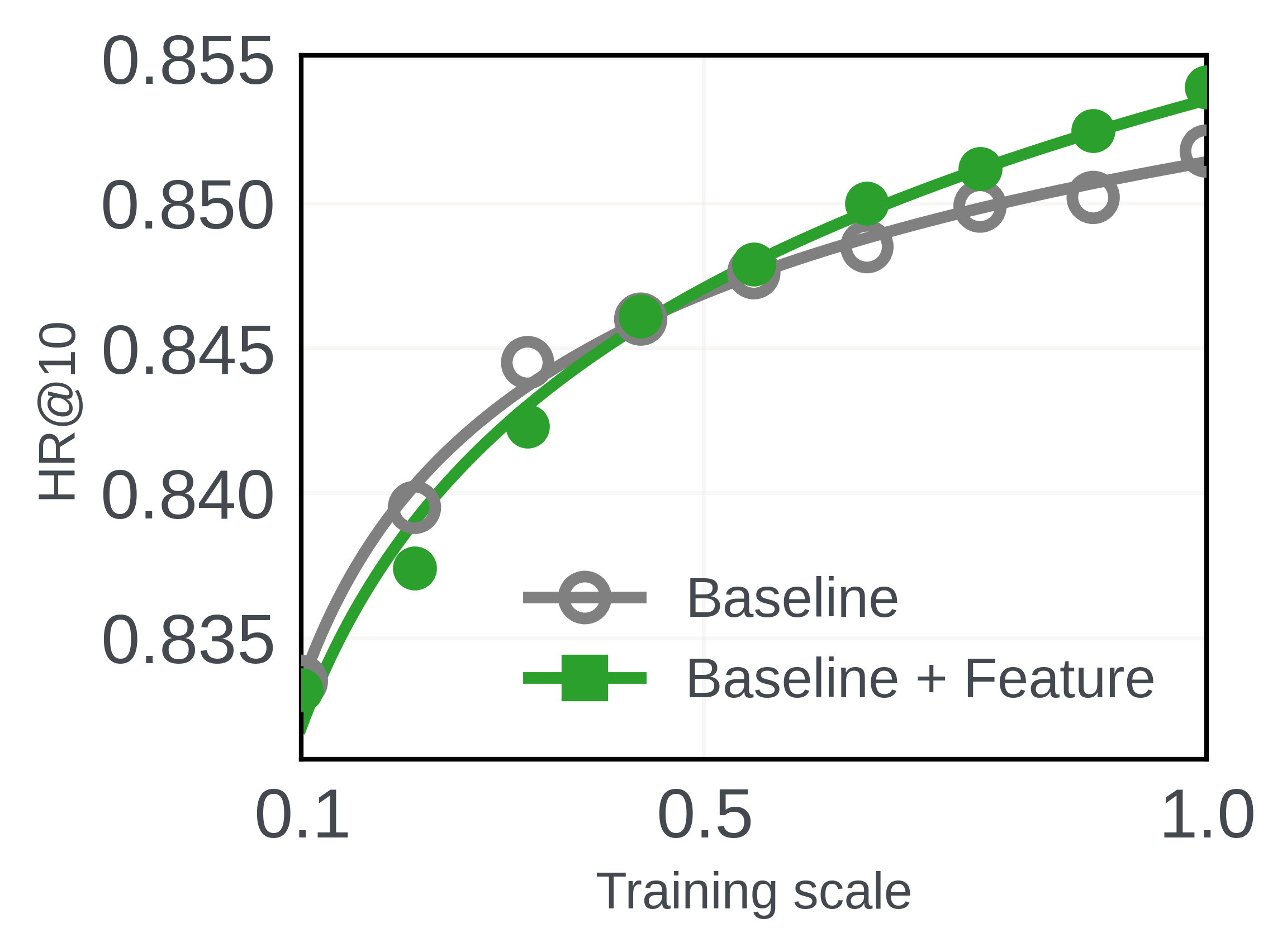}
 \vspace{-5mm}
  \caption{The improvement of a new feature change on the experimental model (green) over the baseline model (grey) only appears at large training scales.}
  \Description{Motivating-example}
  \label{fig:motivating-example}
\end{figure}

We begin with a case study on an LLM-based recommendation model (Fig.~\ref{fig:motivating-example}). We call the original model the \textit{baseline model}, and the model with a new feature added the \textit{experimental model}. In this case study, the new feature change is movie genre information added to the training prompt. We want to know if this feature improves the model after full-scale training. If the performance gain is neutral, we do not want to add the feature, because it increases inference latency. In this particular example, we vary the training scale by changing the amount of training data from MovieLens-20M \cite{movielens}.

\textbf{Misleading signals from small-scale experiments}. We first ask ourselves: How informative is a single small-scale experiment? Fig.~\ref{fig:motivating-example} shows that the HR@10 of the experimental model is lower than that of the baseline model at a small training scale. However, this small-scale result is misleading because genre information actually produces a significant performance improvement at the full training scale. Therefore, relying on the result of a single small-scale experiment here would have led us to mistakenly reject a new feature change that is, in fact, useful for a recommendation model. At first glance, we might be tempted to move the small-scale experiment to the 0.5 scale to produce a more conclusive result than at 0.1 scale. However, we cannot do this in an \textit{ex-ante} (before-the-fact) manner because our experiments (Fig.~\ref{fig:feature-invariant-scaling-wukong-hstu}) show that different categories of features produce different signals across small scales.

\textbf{Scaling law}. Although a single small-scale experiment gives a misleading signal, Fig.~\ref{fig:motivating-example} shows that the \textit{change} in model performance across training scales is nonetheless predictable. In this example, a power law $Loss = aN^bD^c$ \cite{kaplan2020scalinglawsneurallanguage,li2025predictablescaleistep} (where $N$ is model size and $D$ is training data size) fits both the baseline and the experimental model with $R^2>0.99$, each with its own fitted coefficients $a$, $b$, and $c$. This suggests that we can instead run a series of small-scale experiments and extrapolate them to predict the effect of a new feature at full scale. In our paper, we formalize this idea as a statistical inference problem \cite{bishop2006pattern}. This approach is easy to interpret and naturally handles training noise that arises in between training runs.

\section{Feature-invariant scaling}
\begin{figure}[h]
  \centering
 \includegraphics[width=\linewidth]{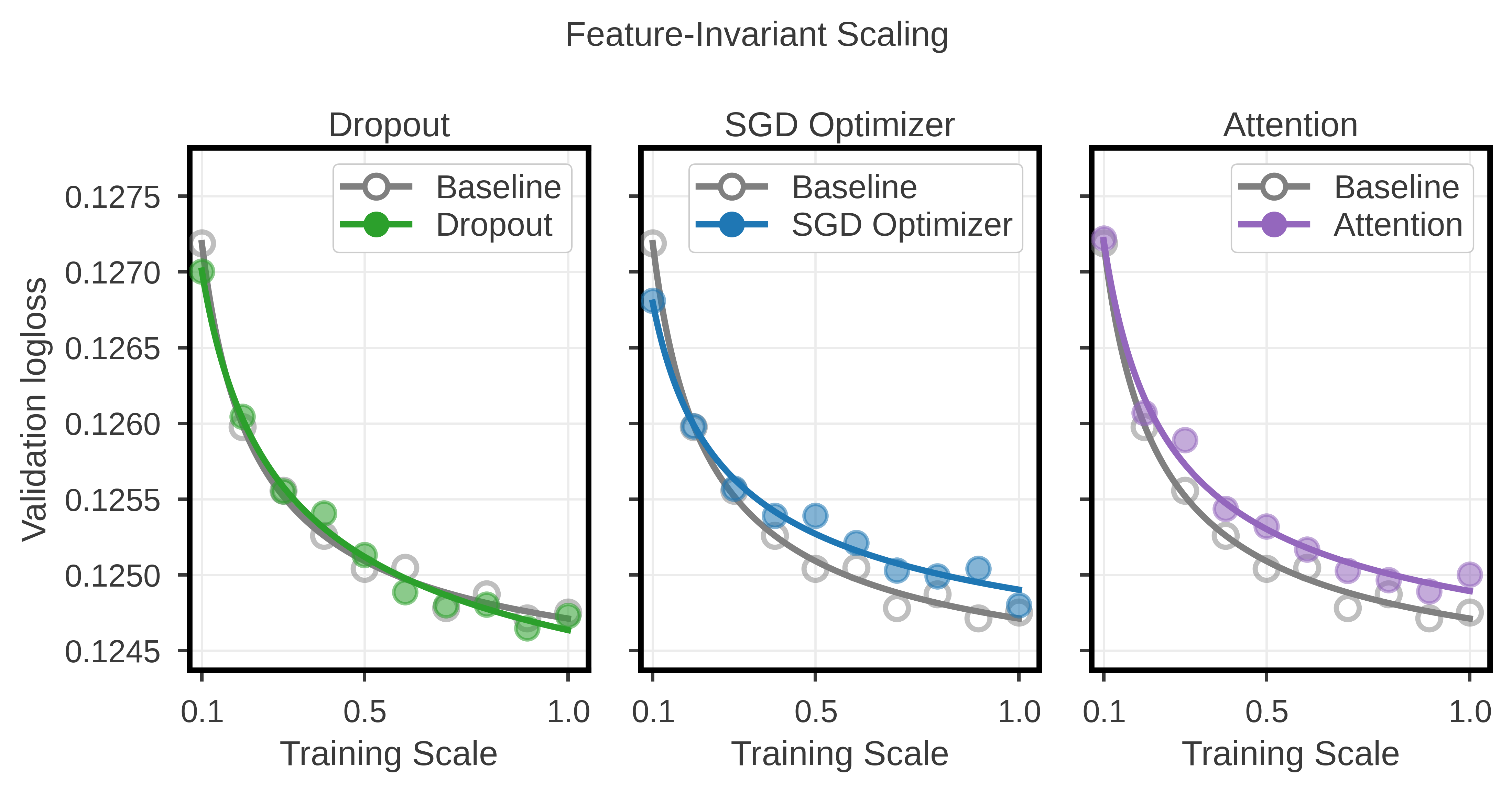}
  \caption{Feature-invariant scaling: Wukong recommendation model \cite{zhang2024wukongscalinglawlargescale} with different features added still follows scaling laws (with different fitted coefficients).}
  \Description{Feature-invariant scaling}
  \label{fig:feature-invariant-scaling}
\end{figure}

Before we can use scaling laws to predict the performance of an experimental model, we need to make sure that the scaling behavior holds across different feature changes. In this section, we show that recommendation models enjoy \textit{feature-invariant scaling}. We ran a small experiment: we applied three different feature changes (adding dropout layer, changing optimizer to SGD, and adding attention layer) to the Wukong recommendation model \cite{zhang2024wukongscalinglawlargescale} and observed the scaling law of the three experimental models (See Fig.~\ref{fig:feature-invariant-scaling}). Our results show that the performance of a recommendation model with different features added is governed by the same scaling law function family with different fitted coefficients. The scaling law fit generally enjoys high $R^2$ (see Table~\ref{tab:feature-invariant-scaling-fit}) regardless of whether a feature is detrimental (e.g., using SGD) or beneficial (e.g., adding dropout) to the model. Specifically, this suggests that when we apply different features to the baseline model, the change in model performance with training scale should still follow a scaling law in the same functional family. We demonstrate that feature-invariant scaling holds true across a large variety of different features and recommendation model families later in Sec.~\ref{subsec:exp-feature-invariant-scaling}.

\begin{table}[h]
  \caption{Scaling law fit with power-law: $Loss=A\,N^{B}D^{C}$
           for the Wukong baseline model with different feature changes applied.}
  \label{tab:feature-invariant-scaling-fit}
  \begin{tabular}{lccc}
    \toprule
    Method &  $R^2$ \\
    \midrule
    Baseline &   0.9949 \\
    Dropout  & 0.9953 \\
    SGD Optimizer  & 0.9923 \\
    Attention  & 0.9876 \\
    \bottomrule
  \end{tabular}
\end{table}

\section{Method}
\label{sec:method}

We assume the performance of the baseline
full-scale recommendation model is known but retraining the model at full
scale for every new feature change is compute intensive. This is a practical assumption in an industrial setting, since performing A/B experimentation with a new feature change requires us to retrain the model with a large amount of data.





\subsection{Problem formulation}

We define $s \in [0,S]$ as the training scale, where $S$ is the
\textit{full-scale}. In recommendation models, the full-scale represents the largest recommendation model (trained on the largest amount of data) that is used during serving time. Specifically, $s$ represents the number of training FLOPs: $s=ND$ \cite{kaplan2020scalinglawsneurallanguage}, where $N$ is the number of training parameters in the recommendation model and $D$ is the training data size (i.e., number of data points). In general, the exact training FLOPs used could be proportional to $s=ND$ by a constant\footnote{For example, in LLM training, the number of training FLOPs is commonly treated as $6ND$ \cite{kaplan2020scalinglawsneurallanguage,hoffmann2022trainingcomputeoptimallargelanguage}.}. However, this constant does not influence the rest of our formulation because it will be absorbed as learnable coefficients (see next section). When scaling the number of model parameters $N$, we keep the model's aspect ratio roughly similar by uniformly scaling the model across sparse feature embedding dimensions, intermediate layer dimensions, and number of model layers. In general, doing so improves training stability and is widely adopted in existing scaling law literature \cite{zhang2024wukongscalinglawlargescale, PLES}.

Similar to Sec.~\ref{sec:motivating-case-study}, we define the \textit{baseline model} as a recommendation model without any added features and the \textit{experimental model} as the baseline model with a new feature added. More formally, let $L_{\mathrm{feature}}(S)$ denote the \textit{unknown}
validation loss of the experimental model at full-scale $S$ and
$L_{\mathrm{baseline}}$ denote the \textit{known} validation loss of the baseline model at full-scale (assumed to be a fixed constant that we want to improve on). For any new feature, we train the experimental model at a collection of smaller training scales
\begin{equation}
    \mathcal{S}
    =
    \{s_1,\ldots,s_n\},
    \qquad
    0 < s_1 < \cdots < s_n < S,
\end{equation}
and observe the experimental model validation losses
$\{y_1,\ldots,y_n\}$ at these small-scales. From here on, we denote these small-scale observations as $\mathcal{D}
    =
    \{(s_i,y_i)\}_{i=1}^{n}$. Each training scale $s_i$ corresponds to a particular model scale $N_i$ and data scale $D_i$.

Although we formulate our problem using validation loss, for which smaller values are
preferred, the same problem setting can be extended to metrics such as AUC or HR@10, as shown in our experiments later.

\subsection{Scaling law with noisy observations}

\textbf{Assumption.}
We first make an assumption to describe how the feature validation loss changes with respect to scale. Consistent with observations from previous sections (Fig.~\ref{fig:feature-invariant-scaling}) and prior work~\cite{kaplan2020scalinglawsneurallanguage, hoffmann2022trainingcomputeoptimallargelanguage}, we assume the validation loss of a recommendation model at training scale $s$ follows a power-law in both model size $N$ and training data size $D$: 
\begin{equation}
    L_{\mathrm{feature}}(s)
    =
    \theta_1 N^{\theta_2} D^{\theta_3},
    \label{eq:power-law}
\end{equation}
where $\theta_1, \theta_2, \theta_3$ are learnable power-law coefficients. Taking the logarithm of both sides yields a linear model:
\begin{equation}
    \log L_{\mathrm{feature}}(s)
    =
    \log \theta_1
    +
    \theta_2 \log N
    +
    \theta_3 \log D.
    \label{eq:log-linear-scaling}
\end{equation}

We assume every small-scale observation $y_i$, obtained at model size $N_i$ and data size $D_i$, is corrupted by a small amount of Gaussian noise $\epsilon_i$, which arises from model training:
\begin{equation}
    y_i
    =
    \log L_{\mathrm{feature}}(s_i)
    +
    \epsilon_i,
    \qquad
    \epsilon_i
    \overset{\mathrm{i.i.d.}}{\sim}
    \mathcal{N}(0,\sigma^2).
    \label{eq:observation-model}
\end{equation}
In general, we observe that the training noise is relatively stationary as we vary the model size. This is because if the models are scaled appropriately with correct normalization and regularization during training, the training variance remains relatively constant. Therefore, our observation of the scaling law can be expressed as a linear regression model:
\begin{equation}
    \mathbf{y}
    =
    \mathbf{X}\boldsymbol{\theta}
    +
    \boldsymbol{\epsilon},
\end{equation}
where
\begin{equation}
    \boldsymbol{\theta}
    =
    \begin{bmatrix}
        \log \theta_1 \\
        \theta_2 \\
        \theta_3
    \end{bmatrix},
    \qquad
    \mathbf{X}
    =
    \begin{bmatrix}
        1 & \log N_1 & \log D_1 \\
        \vdots & \vdots & \vdots \\
        1 & \log N_n & \log D_n
    \end{bmatrix},
    \qquad
    \mathbf{y}
    =
    \begin{bmatrix}
        y_1 \\
        \vdots \\
        y_n
    \end{bmatrix}.
\end{equation}

The parameters $\boldsymbol{\theta}$ and $\sigma$ can be estimated from the small-scale observations in closed form~\cite{gelman1995bayesian}:
\begin{equation}
    \widehat{\boldsymbol{\theta}}
    =
    \begin{bmatrix}
        \widehat{ \log \theta}_1 \\
        \widehat{\theta}_2 \\
        \widehat{\theta}_3
    \end{bmatrix}
    =
    (\mathbf{X}^{\top}\mathbf{X})^{-1}\mathbf{X}^{\top}\mathbf{y},  
    \qquad
    \widehat{\sigma}
    =
    \sqrt{
        \frac{
            (\mathbf{y}-\mathbf{X}\widehat{\boldsymbol{\theta}})^{\top}
            (\mathbf{y}-\mathbf{X}\widehat{\boldsymbol{\theta}})
        }{n-3}
    }.
\end{equation}
Note that we use the term $n-3$ because we have $3$ unknown parameters in Eq.~\ref{eq:power-law}. If we had only scaled model or data size alone, then the denominator becomes $n-2$.

\subsection{Statistical inference}
\label{subsec:bayesian-inf}

Under our model, let $\log \hat{L}_{\mathrm{feature}}(S_*)$ be our prediction of the experimental model's log validation loss at full scale, i.e., at model size $N_*$ and data size $D_*$, conditioned on small-scale observations. The posterior distribution of $\log \hat{L}_{\mathrm{feature}}(S_*)$, conditioned on the small-scale observations $\mathcal{D}$, follows a Student-$t$ distribution~\cite{gelman1995bayesian}: 
\begin{equation}
    \log \hat{L}_{\mathrm{feature}}(S_*) \mid \mathcal{D}
    \sim
    t_{n-3}\!\left(
        \mathbf{x}_{*}^{\top}\widehat{\boldsymbol{\theta}},
        \widehat{\sigma}
        \sqrt{
            \mathbf{x}_{*}^{\top}
            (\mathbf{X}^{\top}\mathbf{X})^{-1}
            \mathbf{x}_{*}
        }
    \right),
    \qquad
    \mathbf{x}_{*}
    =
    \begin{bmatrix}
        1 \\
        \log N_* \\
        \log D_*
    \end{bmatrix}.
    \label{eq:full-scale-posterior}
\end{equation}

Note that without any prior knowledge of parameters, the formulation above is consistent for both the Bayesian \cite{gelman1995bayesian} and Frequentist \cite{montgomery2022introduction} framework and in both cases the prediction interval of the full-scale experimental performance is the same (our overall method does not change significantly in either the Bayesian or Frequentist framework). Since the baseline model loss at full scale
$L_{\mathrm{baseline}}$ is a constant that we want to outperform, the probability
\begin{equation}
    \Pr\!\left(
        \log \hat{L}_{\mathrm{feature}}(S_*)
        <
        \log L_{\mathrm{baseline}}
        \mid \mathcal{D}
    \right)
    \label{eq:posterior-probability-less-than-baseline}
\end{equation}
represents the probability (conditioned on the small-scale observations) that the experimental model will achieve a lower validation loss than
the baseline model at full scale.

\begin{figure}[h]
  \centering
 \includegraphics[width=\linewidth]{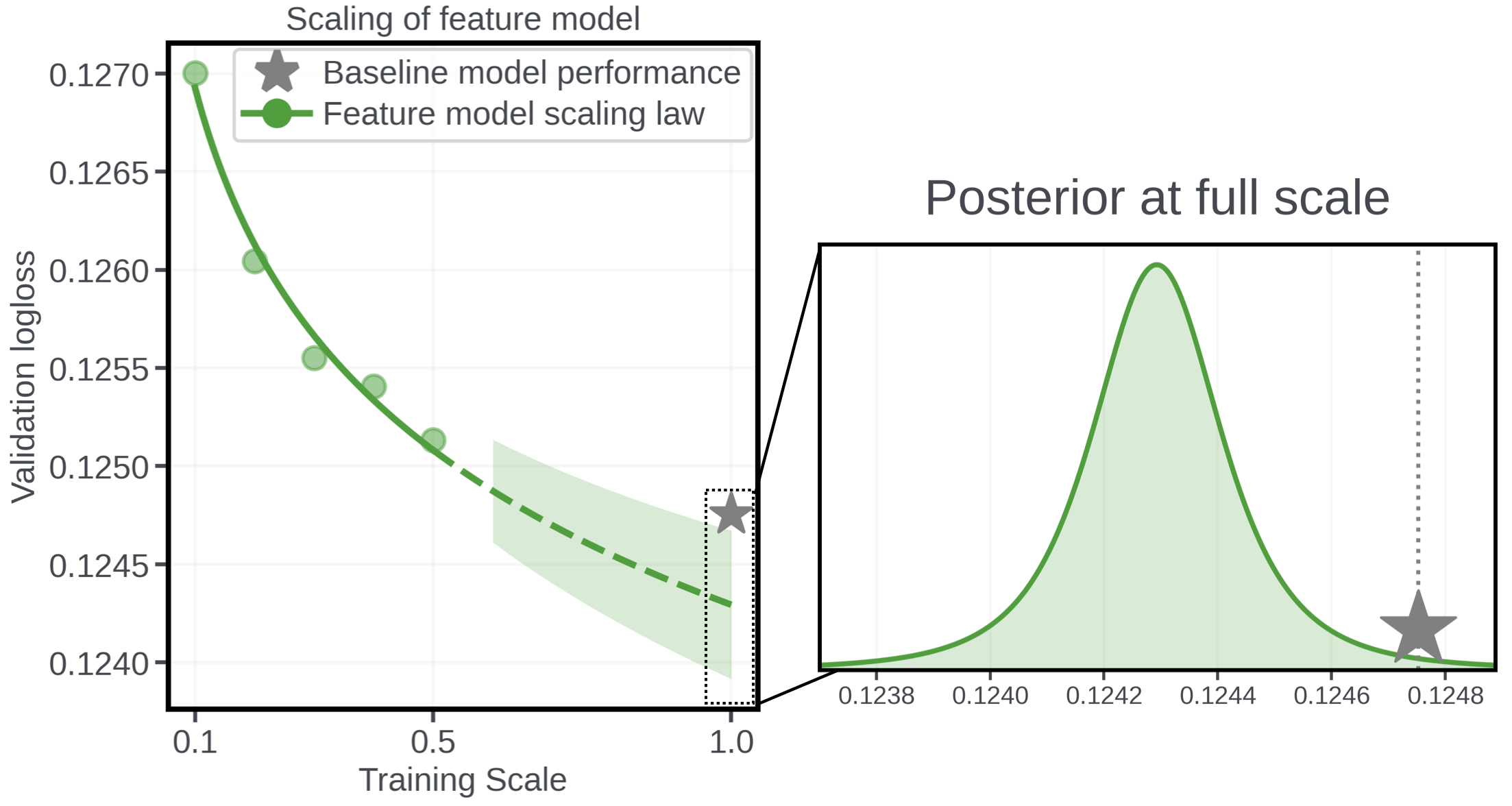}
  \caption{The fitted scaling law (left) produces the probability distribution of the experimental model performance at full scale $S_*=(N_*, D_*)$ (right), allowing us to predict the probability of improvement of a new feature over the baseline model.}
  \Description{posterior-distribution-image}
  \label{fig:posterior-full-scale}
\end{figure}

An example of the distribution of the performance of an experimental model at full scale is shown in Fig.~\ref{fig:posterior-full-scale}, indicated by the histogram on the right. Theoretically, our plot is three dimensional because we have two scaling factors (model and data size). However, we map them into a single training scale to visualize the scaling curve in one dimension. In this example, the baseline model performance at full scale is the \textcolor{gray}{gray star} and there is a high probability of the experimental model outperforming the baseline model at full scale.


\section{Resource savings from small-scale experiments}

A natural question is whether a sequence of small-scale experiments is actually less resource-intensive than a single full-scale training run. If we co-scale both training and model scale, a half-scale run uses approximately one quarter of the cost of a full-scale run. Consequently, a sweep of small-scale experiments at $s \in \{0.1, 0.2, 0.3, 0.4, 0.5\}$ incurs about $55\%$ of the resource needed to run a single full-scale training run.

In addition, co-scaling data and model size offers another practical and important benefit. In many industrial settings, experimentation is often \textit{GPU-bound}. That is, the compute available for experimentation may simply be insufficient to hold the full-scale model in memory. In such settings, a full-scale run to validate a new feature change on the full-sized baseline model is simply impossible. In this case, down-scaling model size in small-scale experiments is the only way for us to perform A/B experiments.

\section{Practical Considerations}
\label{sec:practical-consideration}

Next, we discuss three practical considerations that practitioners and researchers should have when exploiting scaling laws with our method.

\textbf{Choice of scaling axis.}
In our formulation, we set $s=ND$. The main advantage of jointly reducing the model size $N$ and the amount of training data $D$ is the substantially greater resource savings from smaller experimental models. The theoretical formulation of our method does not change if we choose to only adjust one scaling axis. For instance, if we fix the model scale $N$ and only vary training data scale $D$, the fixed model scale $N$ can be factored out in Eq.~\ref{eq:power-law} as a constant. In our paper, we also vary the model size by scaling the number of layers, width, and size of embedding tables fairly uniformly (see Sec.~\ref{sec:setup-experiments} for more details). We believe that our formulation and method will still work in specialized settings such as scaling of sparse activation in Mixture-of-Experts \cite{shazeer2017outrageouslylargeneuralnetworks}.

\textbf{Decision rule.}
Our method (Sec.~\ref{sec:method}) estimates the posterior probability that an experimental model outperforms the baseline model at full-scale. However, in industrial applications, a marginal improvement\footnote{Here, a marginal improvement refers to the case where, in a noiseless setting, a feature actually improves a baseline model but the amount of improvement is relatively small.} might not justify the deployment of a new feature, especially if it introduces additional computation, storage, and engineering complexity. It is therefore more practically useful to determine whether the feature provides an improvement larger than a meaningful threshold (e.g., 1\% better than the baseline) at full-scale. For instance, we can replace the original comparison in Eq.~\ref{eq:posterior-probability-less-than-baseline} with
\begin{equation}
    \Pr\!\left(
        \hat{L}_{\mathrm{feature}}(S)
        <
        (1-\delta)L_{\mathrm{baseline}}
        \mid \mathcal{D}
    \right)
    \label{eq:significant-improvement}
\end{equation}
to predict if a new feature reduces the validation loss of the baseline model by more than $\delta$ percentage points at full-scale. A Bayesian decision rule could be that if this probability exceeds 90\%, then we consider the feature effective for the baseline model (and we can proceed with higher-fidelity validation). In industrial settings, a 0.1\% reduction in loss typically leads to more than 0.5\% increase in topline metrics across billions of users \cite{zhai2024actionsspeaklouderwordsHSTU}.

\textbf{Alternative functional forms.}
Although power laws provide an empirically effective scaling law function in many settings, some work has identified scaling relationships that are better described by other functional forms~\cite{muennighoff2025scalingdataconstrainedlanguagemodels}. Our paper's general approach does not necessarily require scaling laws to follow power-laws. The key difference is that for nonlinear scaling laws, the posterior distribution of the parameters and the full-scale performance may no longer be available in closed form. In such cases, it can be approximated using standard Bayesian inference techniques, such as MCMC \cite{speagle2020conceptualintroductionmarkovchain}. We provide the specific mathematical formulation for the nonlinear scaling law in App.~\ref{app:nonlinear}.

\begin{figure*}[t] \centering \includegraphics[width=0.9\textwidth]{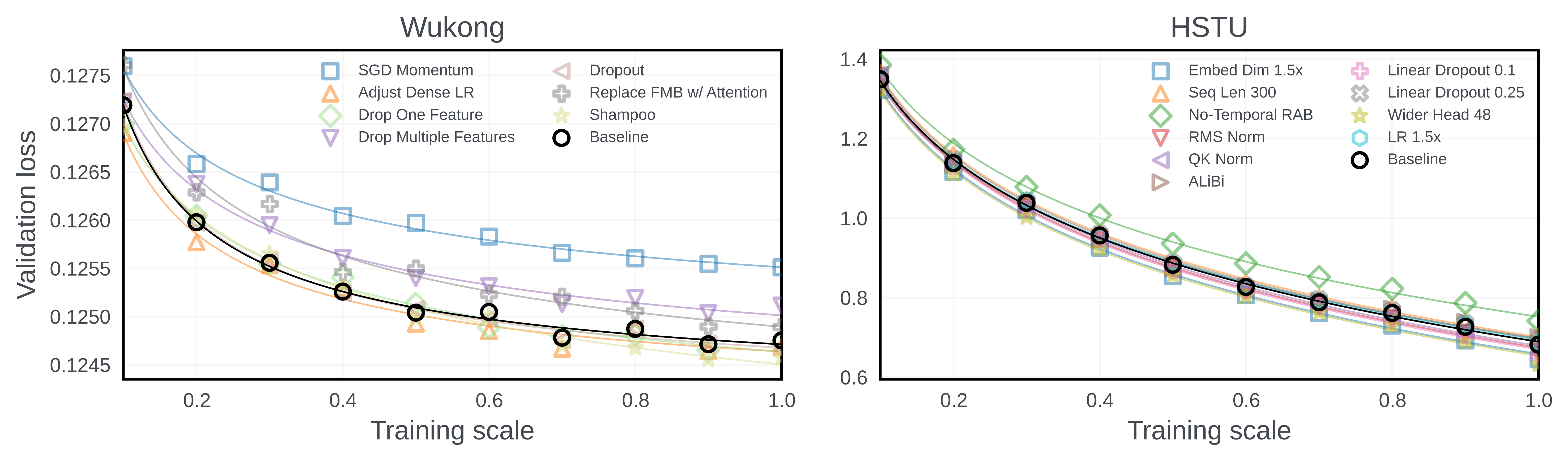}
\caption{Feature-invariant scaling laws for Wukong \cite{zhang2024wukongscalinglawlargescale} and HSTU \cite{zhai2024actionsspeaklouderwordsHSTU} over a wide variety of model, data, and optimizer features added. Regardless whether a feature improves or regresses the model, the model validation loss changes predictably with training scale.}
\label{fig:feature-invariant-scaling-wukong-hstu}
\end{figure*}
\section{Experiments}
In this section, we perform extensive experiments on three families of recommendation models---Wukong \cite{zhang2024wukongscalinglawlargescale}, HSTU \cite{zhai2024actionsspeaklouderwordsHSTU}, and an LLM-based recommendation model \cite{liang2026gr2generativereasoningreranker,Zhao_2024_recsys_llm2}. Our results are divided into three sections.

\begin{enumerate}
    \item We demonstrate that all three families of recommendation models exhibit feature-invariant scaling laws.
    \item We demonstrate the effectiveness of our method in predicting whether a new feature will improve model performance from a series of small-scale experiments. In particular, we show that we can predict accurately and with high confidence the effectiveness of a new feature at a fraction of a full training run.
    \item We present several interesting ablations and case studies, including the prevalence of the ``crossover'' effect.
\end{enumerate}

\subsection{Experimental Setup}\label{sec:setup-experiments}

\textbf{Wukong}. Our recommendation task is to make click-through-rate prediction for the Criteo dataset. At full-scale, the training data contains
  $\sim\!4.2\mathrm{B}$ data points (23 days of data) and the model is evaluated on the last day of data, containing $\sim\!178\mathrm{M}$ data points. We use a Wukong model with the following configuration: the number of Linear Compress Block (LCB)
embeddings $n_L=16$, number of Factorization-Machine Block (FMB) embeddings $n_F=16$, number of compressed embeddings $k=20$, an MLP size of $2 \times 1024$, number of Wukong layers $l=5$. We vary the training scale $s \in [0.1, 1.0]$ by uniformly sampling the
  training data and adjusting the model embedding dimension $d$, scaled from $8$ at
  $s{=}0.1$ to $64$ at $s{=}1.0$. Changing $d$ uniformly adjusts the sparse embedding table width, the dense-feature projection dimension, and the FMB and LCB output
  projection dimensions.

\noindent \textbf{HSTU}. Our task is to predict the next movie rated by a user based on their past movie interactions from the \textbf{MovieLens-20M} dataset. We treat the last movie rated by each user in the dataset as the validation dataset. We use the HSTU-large model as described in \cite{zhai2024actionsspeaklouderwordsHSTU}. We co-scale three architectural dimensions --- the item-embedding
(hidden) dimension $d$, the number of transformer
blocks $L$, and the number of attention heads
$H$. At $s=1.0$, the model reaches the HSTU-large configuration
($d{=}256$, $L{=}16$, $H{=}8$). At $s=0.1$, the model shrinks to
($d{=}26$, $L{=}2$, $H{=}1$).

\noindent \textbf{LLM-based recommendation system}. We fine-tune Llama-3.1-Instruct \cite{touvron2023llamaopenefficientfoundation} from its pre-trained checkpoint to predict the next movie a user will rate based on their historical interaction with other movies. In this setting, we choose to fix the model size and only vary the training data tokens. We give more details and examples on the training data format in Fig.~\ref{fig:LLM-setup}.

\noindent \textbf{Training setup}. Similar to the original paper, we trained Wukong for $1$ epoch on 8 A100 GPUs with a per-GPU batch-size of $8196$, using a row-wise
Adagrad optimizer for the sparse feature table with learning rate $4.0\times10^{-2}$ and an AdamW optimizer with learning rate $3.5\times10^{-3}$. We trained HSTU for
$100$ epochs with the AdamW optimizer at learning rate $1\times10^{-3}$. We perform full-parameter fine-tuning of the LLM with 8 A100 GPUs for 1 epoch with a per-GPU batch-size of $2$, using AdamW with $\beta=(0.9,0.999)$ and a learning rate of $5\times10^{-6}$. If some feature changes increase FLOPs, the experimental model full-scale has the same FLOPs as the baseline.

\noindent \textbf{Evaluation metrics.}
Throughout our experiments, we fit the scaling-law curve to the validation loss of the recommendation models. However, we noticed that the scaling behavior for recommendation models still holds true for other evaluation metrics such as HR@10 (See Sec.~\ref{sec:results-llm-recsys}). This strongly suggests that our method's effectiveness holds true even for other evaluation metrics found throughout recommendation system literature.

\subsection{Feature-invariant Scaling Laws for Recommendation Models}\label{subsec:exp-feature-invariant-scaling}

We consider a broad variety of features spanning optimizer, model, and data changes applied to recommendation models. We have more than 40 features in total. However, due to space constraints, we only present a subset of features in our main paper. The complete description and results for all features can be found in App.~\ref{app:all-feature-description} and App.~\ref{app:all-result}.

We first demonstrate that recommendation models enjoy feature-invariant scaling across model families and different categories of features. We co-scale the training data and model size for the baseline model with a variety of features added to it and observe how the model validation loss changes with training scale. Fig.~\ref{fig:feature-invariant-scaling-wukong-hstu} shows that even though a new feature could improve or regress the overall performance of a recommendation model, the model performance can still be captured by a power law with varying coefficients. In fact, we observe that every scaling curve can be fitted with a power-law to an extremely high $R^2$ value (See Table~\ref{tab:feature-scaling-r-square-fit}). This suggests that recommendation models do indeed enjoy \textit{feature-invariant scaling}.

\subsection{Exploiting scaling to predict effectiveness of a new feature}

\suppressfloats[t]
\begin{table}[t]
  \centering
  \footnotesize
  \setlength{\tabcolsep}{3pt}
  \caption{Scaling law (power-law) fit of different new feature changes applied to different model families.}
  \label{tab:feature-scaling-r-square-fit}

  \begin{minipage}[c]{0.47\columnwidth}
    \centering
    \begin{tabular}{lc}
      \toprule
      \textbf{Wukong}           & $R^2$ \\
      \midrule
      Baseline                  & 0.9949 \\
      SGD Momentum              & 0.9923 \\
      Adjust Dense LR           & 0.9803 \\
      Drop One Feature          & 0.9927 \\
      Drop Multiple Features    & 0.9920 \\
      Dropout                   & 0.9957 \\
      Attention                 & 0.9807 \\
      Shampoo                   & 0.9934 \\
      \bottomrule
    \end{tabular}

    \vspace{2mm}

    \begin{tabular}{lc}
      \toprule
      \textbf{LLM (Llama-3.1-8B)} & $R^2$ \\
      \midrule
      Baseline          & 0.9977 \\
      Genre Information & 0.9970 \\
      No Rating         & 0.9981 \\
      Dropout           & 0.9939 \\
      Extra Layer       & 0.9743 \\
      \bottomrule
    \end{tabular}
  \end{minipage}
  \hfill
  \begin{minipage}[c]{0.47\columnwidth}
    \centering
    \begin{tabular}{lc}
      \toprule
      \textbf{HSTU}         & $R^2$ \\
      \midrule
      Baseline              & 0.9989 \\
      Embed Dim 1.5x        & 0.9988 \\
      Seq Len 300           & 0.9989 \\
      No-Temporal RAB       & 0.9981 \\
      RMS Norm              & 0.9988 \\
      QK Norm               & 0.9984 \\
      ALiBi                 & 0.9972 \\
      Linear Dropout 0.1    & 0.9984 \\
      Linear Dropout 0.25   & 0.9988 \\
      Wider Head 48         & 0.9982 \\
      LR 1.5x               & 0.9987 \\
      \bottomrule
    \end{tabular}
  \end{minipage}
\end{table}

In this section, we present the main results associated with our method which \textit{exploits feature-invariant scaling laws} (Sec.~\ref{sec:method}) to predict if a feature is useful for a baseline recommendation model. Using our method, we were able to evaluate more than 40 new data and model changes and uncover a few that improve upon HSTU's state-of-the-art performance on the MovieLens-20M dataset. We construct the following setup: for each new feature, we train and observe the model validation loss at scaling intervals $s \in$ \{0.1,0.2,0.3,0.4, 0.5\}. Then, we fit these short-horizon performance observations to a scaling law (whose functional form follows a power-law). We then use the scaling law to predict the experimental model's performance at the full $s=1.0$ scale with an uncertainty interval from Eq.~\ref{eq:full-scale-posterior}.

\textbf{Visualizing uncertainty of full-scale predictions}. We show our method's prediction of full-scale model performance from the effect of adding dropout \cite{hinton2012dropout} to the MLP layers in Wukong and HSTU. Due to space constraints, the prediction plots for other features can be found in Fig.~\ref{fig:extrapolation_all_features} in the appendix. Fig.~\ref{fig:method-prediction-with-confidence} visually shows that, conditioned on small-scale observations, the uncertainty of our prediction (represented by the confidence intervals) of the performance of the full-scale experimental model decreases with more small-scale experiments. This uncertainty is computed in closed-form according to Eq.~\ref{eq:full-scale-posterior}.

\begin{figure*}[t]
  \begin{minipage}[t]{0.48\textwidth}
    \centering
    \includegraphics[width=0.48\linewidth]{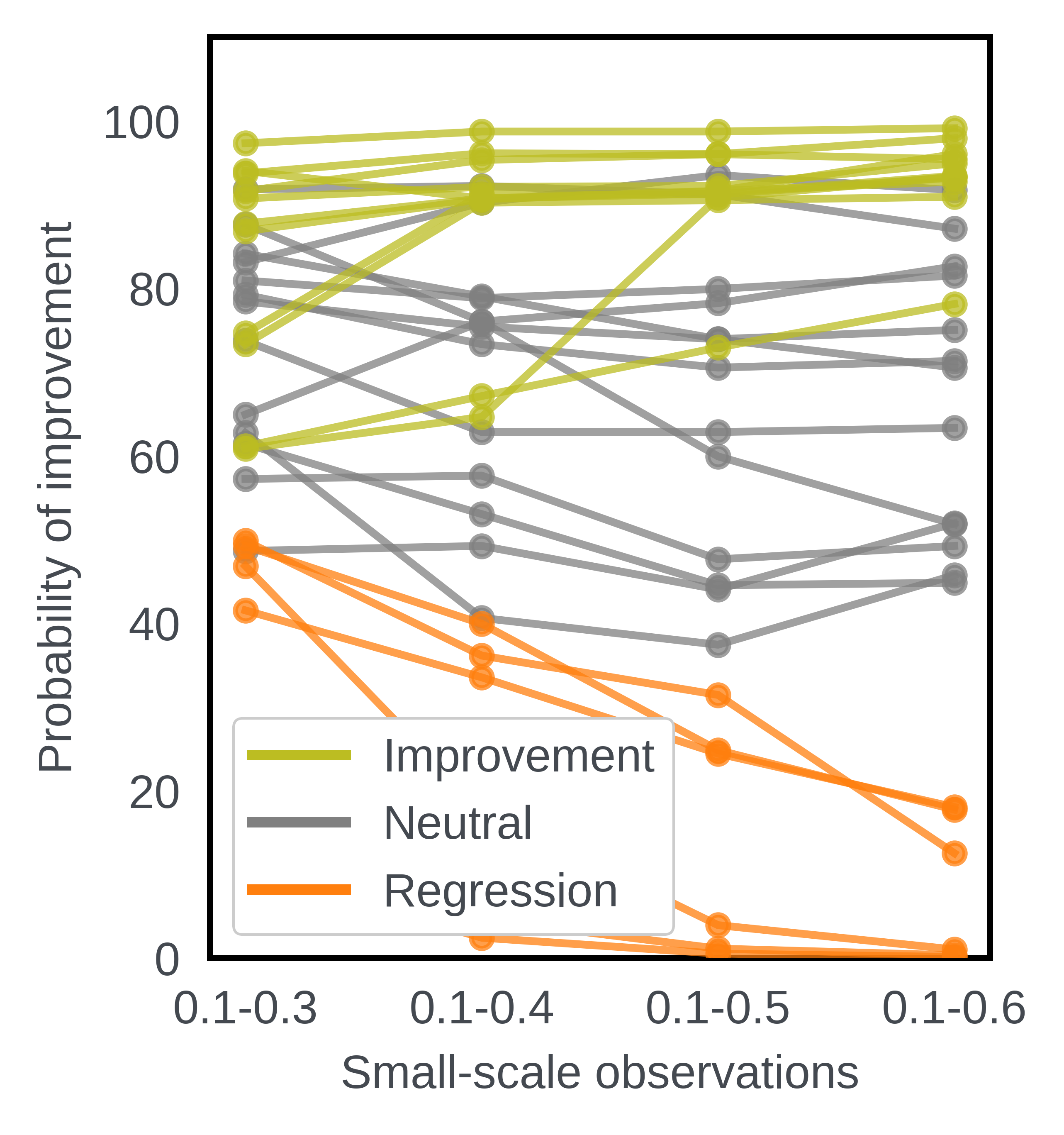} \hfill \includegraphics[width=0.48\linewidth]{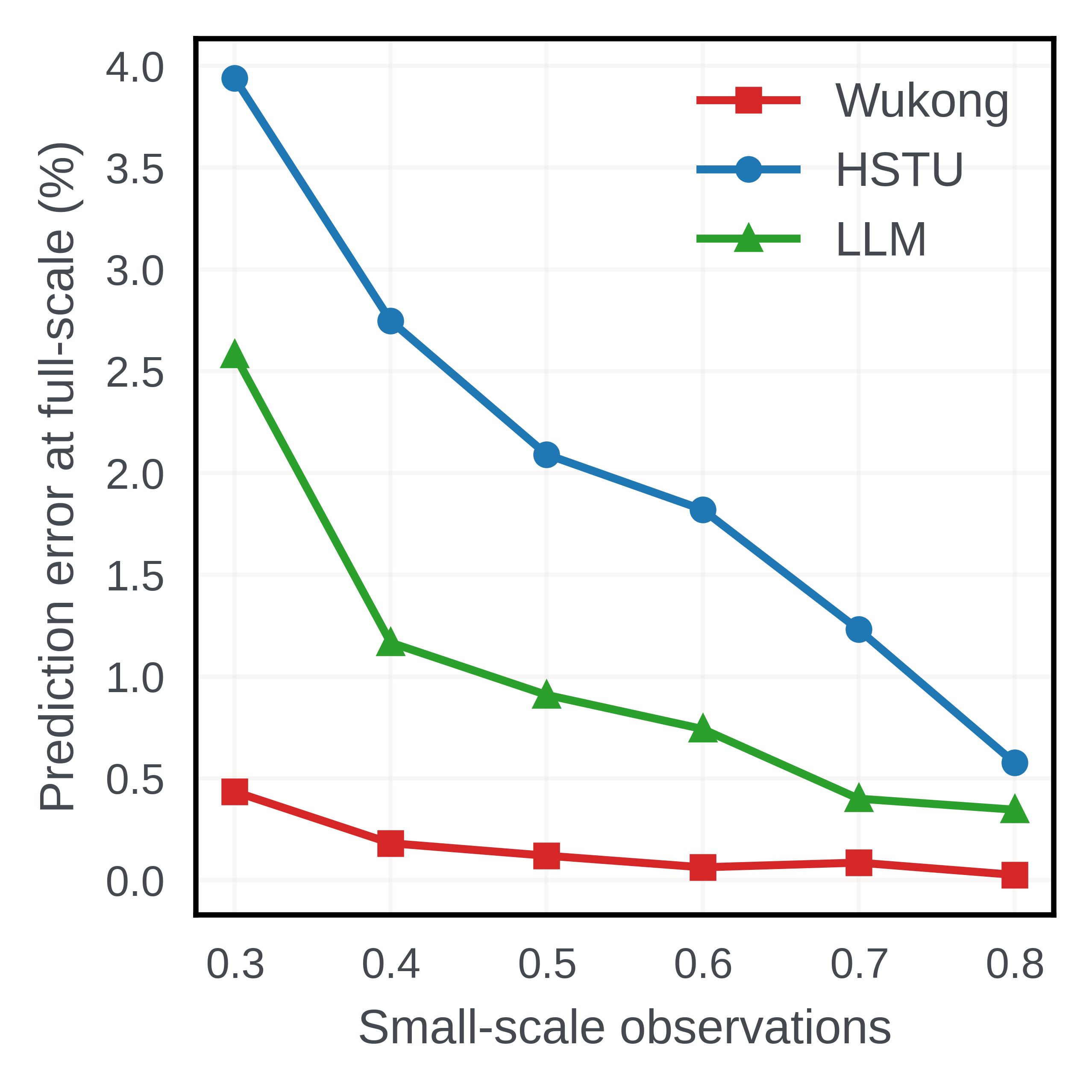}
    \captionof{figure}{\textbf{Left}: The posterior probability of a new feature improving the baseline model (using our method) as we increase the number of small-scale observations. \textbf{Right}: The out-of-sample prediction error of our method of the full-scale performance as we increase the number of small-scale observations.}
    \label{fig:prediction-certainty-error-with-more-scales}
  \end{minipage}\hfill
  \begin{minipage}[t]{0.48\textwidth}
    \centering
    \includegraphics[width=1.0\linewidth]{images/scaling_law/wukong_hstu_prediction.png}
    \captionof{figure}{Increasing the number of small-scale experiments decreases the uncertainty of improvement.}
    \vspace{-10mm}\label{fig:method-prediction-with-confidence}
  \end{minipage}
  \vspace{-5mm}
\end{figure*}
\textbf{Prediction results}. Next, we present results showing that our method makes the correct judgment of the effectiveness of new features. We consider a feature beneficial if it reduces the validation loss of the baseline model by more than 1\% at full scale, i.e., $L_{\mathrm{feature}}(S)
        <
        0.99L_{\mathrm{baseline}}$.
Our Bayesian decision rule is to consider a new feature beneficial if the probability that the experimental model outperforms the baseline model at full-scale by 1\% is more than 90\%, i.e., $\Pr\!\left(
        L_{\mathrm{feature}}(S)
        <
        0.99L_{\mathrm{baseline}}
        \mid \mathcal{D}
    \right) > 0.9$.

\begin{table}[t]
\centering
\caption{Examples of beneficial feature changes applied to HSTU-Large \cite{zhai2024actionsspeaklouderwordsHSTU} improve over the SOTA performance for MovieLens-20M.}
\label{tab:hstu-ml20m-hit-rates}
\small
\begin{tabular}{lccc}
\toprule
\textbf{Feature}
& \textbf{HR@10}
& \textbf{HR@50}
& \textbf{HR@200} \\
\midrule
Baseline \cite{zhai2024actionsspeaklouderwordsHSTU}
& 0.3567
& 0.6139
& 0.8073 \\
\midrule
Time Gap Bias
& 0.3603 $(+1.01\%)$
& 0.6186 $(+0.77\%)$
& 0.8112 $(+0.48\%)$ \\

Dropout 0.1
& 0.3590 $(+0.64\%)$
& 0.6183 $(+0.72\%)$
& 0.8107 $(+0.42\%)$ \\

Pop Debias
& 0.3585 $(+0.50\%)$
& 0.6182 $(+0.70\%)$
& 0.8102 $(+0.36\%)$ \\

Wider Head 48
& 0.3584 $(+0.48\%)$
& 0.6176 $(+0.60\%)$
& 0.8097 $(+0.30\%)$ \\

RMS-Norm
& 0.3584 $(+0.48\%)$
& 0.6175 $(+0.59\%)$
& 0.8094 $(+0.26\%)$ \\

Attn Residue
& 0.3581 $(+0.39\%)$
& 0.6164 $(+0.41\%)$
& 0.8086 $(+0.16\%)$ \\
\bottomrule
\end{tabular}
\end{table}
Each line plot on the left of Fig.~\ref{fig:prediction-certainty-error-with-more-scales} corresponds to the probability of a new feature improving the baseline HSTU (the full list of features can be found in App.~\ref{app:all-feature-description}). In general, for features that actually improve the model, their posterior probability of improvement increases to 100\% with more small-scale experiments. On the other hand, this probability approaches zero for features that regress the model. This strongly suggests that our method and the Bayesian decision rule are effective in determining, at a small-scale, whether a new feature is effective, without actually performing full-scale experimentation, hence saving computational resources.

Last but not least, the right of Fig.~\ref{fig:prediction-certainty-error-with-more-scales} shows that our method produces a small prediction error for the out-of-sample full-scale experimental model performance based on small-scale experiments. In addition, this prediction error decreases with more small-scale observations.

\section{LLM-based Recommendation Models}\label{sec:llm-recsys}

The popularity of \textit{LLM-based recommendation models} \cite{liang2026gr2generativereasoningreranker,xue2026generativerecommendationlargescaleadvertising,xie2026agentictaggerstructureditemrepresentation} has risen in recent years. In this section, we demonstrate that scaling laws can also be exploited in LLM-based recommendation models to predict the effectiveness of a new feature.

We consider a text-based recommendation setup where an LLM (or agent) receives information related to historical interaction events of a user with past items. The LLM is then prompted to recommend a single item from a large pool of candidate items. We probe the probability logit output of the LLM to compute the top $K$ recommended items, which allows us to compute popular evaluation metrics such as HR@K.

\begin{figure}[h]
  \centering
 \includegraphics[width=1.0\linewidth]{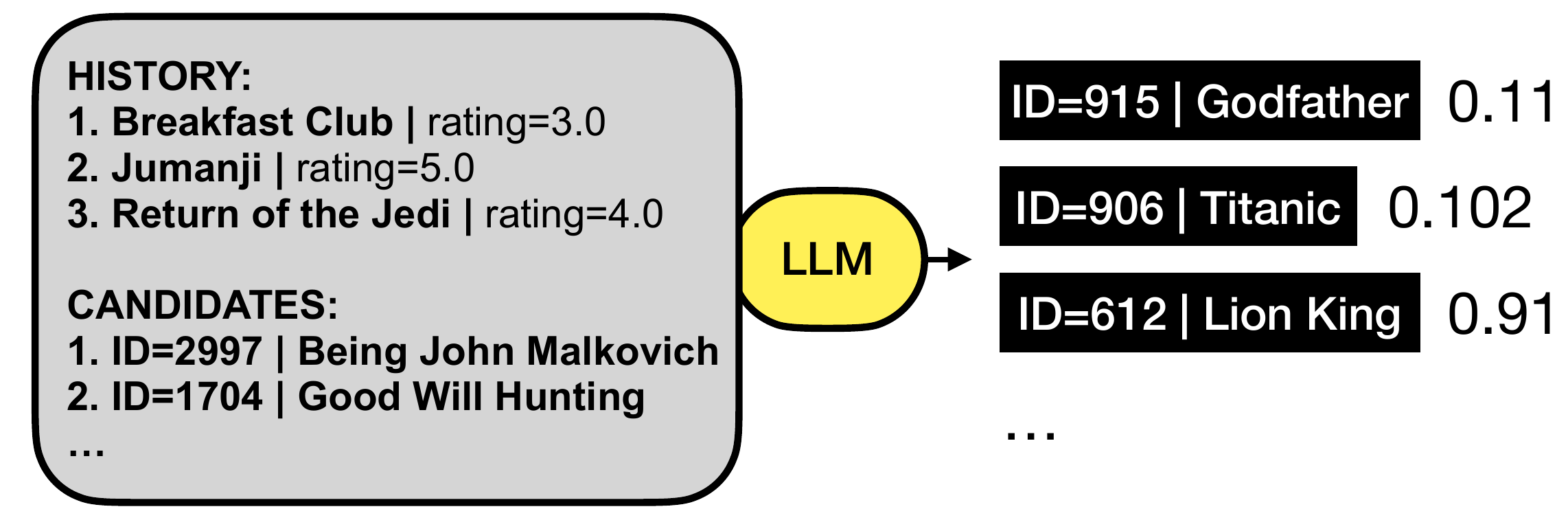}
  \caption{LLM-based recommendation model used in our experiment, shown with the MovieLens-20M dataset.}
  \Description{xxx}
  \label{fig:LLM-setup}
\end{figure}

We perform full-parameter finetuning on pretrained LLMs. Before finetuning, we preprocess the MovieLens-20M dataset so that in each data point, the LLM is given a historical context of the movies that the user has previously watched and rated. A list of 100 candidate movies with the correct next-watched movie inserted randomly is given to the LLM and the LLM is prompted to produce the next movie watched. Fig.~\ref{fig:LLM-setup} shows an example of one such preprocessed data point. The numerical score of each recommended movie corresponds to the probability logits produced by the finetuned LLM. Similar to the HSTU case, we treat the last movie watched by all users as a held-out validation set. Because our candidate set in the prompt is a list of 100 movies, we expect the HR@10 for this setup to be higher than that of HSTU (which considers all movies as candidates).

\vspace{-5mm}
\subsection{Experimental setup}
We performed our experiments across two open-source LLM families---Qwen3-Instruct \cite{yang2025qwen3technicalreport} and Llama-3.1-Instruct \cite{touvron2023llamaopenefficientfoundation}. Because we finetune the model from pretrained checkpoints, we only adjust the training data scale in our experiments. However, we expect our results to hold true even if we co-scale both data and model sizes for LLMs. We finetune the LLMs with 8 A100 GPUs, with an AdamW optimizer and a peak learning rate of $5\times10^{-6}$ with a cosine decay schedule and a 3\% linear warmup, and global gradient-norm clipping at 1.0. We use a batch size of 8 sequences, a maximum sequence length of 4096 tokens, and train for a single epoch.

\subsection{Feature-invariant Scaling Laws for LLM-based Recommendation Models}\label{sec:results-llm-recsys}

We apply four different features during the fine-tuning of Llama-3.1-8B-Instruct. The first two are addition of \textbf{genre} and \textbf{rating} information to each movie in the prompt, while \textbf{dropout} and \textbf{extra layer} are model-specific feature changes to the LLM itself. Detailed descriptions of these features are provided in App.~\ref{app:all-feature-description}. As shown in Fig.~\ref{fig:LLM-7b-8b-scaling}, LLM-based recommendation models also exhibit feature-invariant scaling, just like traditional recommendation models shown earlier. This suggests that our method is equally effective in predicting the effectiveness of new features for LLM-based recommendation systems (as shown in Fig.~\ref{fig:prediction-certainty-error-with-more-scales}). The $R^2$ values reported in each panel are also consistently close to 1, indicating a strong fit to the scaling-law predictions regardless of which feature is applied to the baseline model. Interestingly, these scaling relationships hold not only for validation loss (top row), but also for downstream performance metrics such as HR@10 (bottom row). These results suggest that scaling laws can even be used to predict other performance metrics (e.g., NDCG \cite{jarvelin2017ir}, Normalized Entropy \cite{he2014practical}).

\begin{figure}[h]
  \centering
 \includegraphics[width=1.0\linewidth]{images/LLM_results/llama_8B_horizontal.png}
  \Description{xxx}
  \includegraphics[width=1.0\linewidth]{images/LLM_results/qwen_7B_horizontal.png}
  \caption{Feature-invariant scaling laws for \texttt{Llama-3.1-8B-Instruct} (top) and \texttt{Qwen2.5-7B} (bottom).}
  \Description{xxx}
  \label{fig:LLM-7b-8b-scaling}
\end{figure}


\subsection{Ablation for different LLMs}\label{sec:results-llm-recsys-ablation-model}

\begin{figure*}[t]
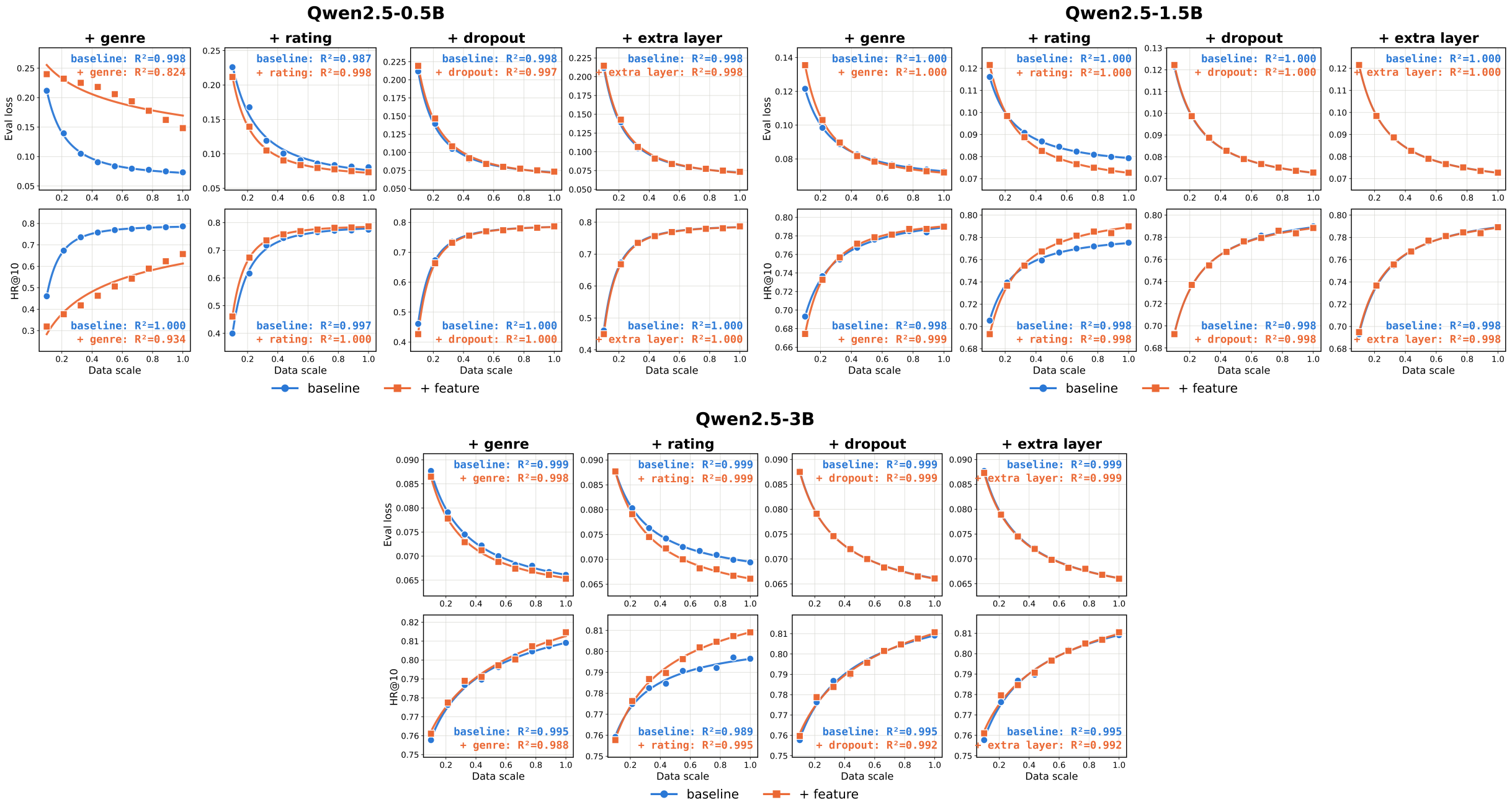

  \centering
 \includegraphics[width=0.45\linewidth]{images/prediction_error_percentage/q05b_features_2x4.png}
  \Description{xxx}
  \includegraphics[width=0.45\linewidth]{images/prediction_error_percentage/q15b_features_2x4.png}
  \includegraphics[width=0.45\linewidth]{images/prediction_error_percentage/q3b_features_2x4.png}
  \caption{Feature-invariant scaling laws for LLMs of different sizes in the Qwen family.}
  \Description{xxx}
  \label{fig:LLM-qwen-sizes}
\end{figure*}

We also found that LLMs across different model families (Qwen and Llama) and sizes (from 0.5B to 8B) consistently (apart from an edge case, which we discuss in Sec.~\ref{subsec:failure-cases}) enjoy feature-invariant scaling. For instance, Fig.~\ref{fig:LLM-7b-8b-scaling} shows that both Qwen-2.5-7B-Instruct and Llama-3.1-8B-Instruct enjoy feature-invariant scaling laws. In addition, the same scaling behavior holds true for LLMs with varying sizes (Fig.~\ref{fig:LLM-qwen-sizes}).

\section{Interesting case studies}\label{sec:interesting-case-study}
In this section, we present some interesting case studies we found in our experiments. The first case study highlights the prevalence of ``crossover'' effects during the training of recommendation models. The second case study presents some edge examples that we encountered in which scaling laws fail.

\subsection{``Crossover points'' in scaling curves for LLM-based recommendation models}\label{subsec:crossover}

\begin{figure}[h]
  \centering
 \includegraphics[width=1.0\linewidth]{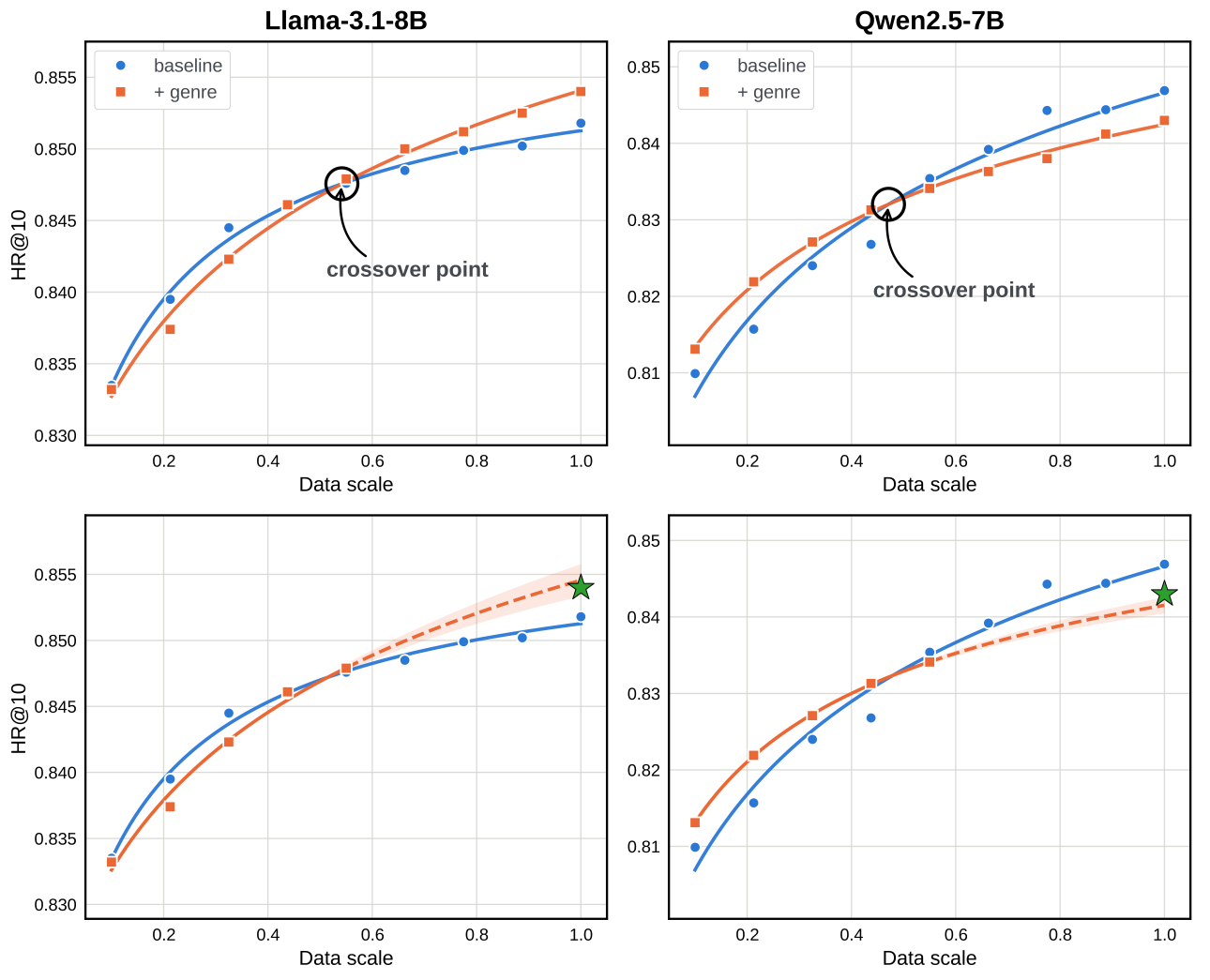}
  \caption{Crossover points are points on the training scales where the signal from a new feature change flips.}
  \Description{xxx}
  \label{fig:cross-over}
\end{figure}

When applying new features to baseline recommendation models, we often observe ``crossover points'' in scaling curves. This is shown in the top row of Fig.~\ref{fig:cross-over}, where the baseline model outperforms (or underperforms) the experimental model at a small-scale but the trend reverses at a larger scale. However, as seen from the bottom row, our method is still effective in predicting the full-scale performance correctly.

In addition, the top row of Fig.~\ref{fig:cross-over} shows that a new feature that is effective for one LLM might not be effective for another LLM family. While adding genre information improves recommendation models powered by Llama, it regresses performance for Qwen. Nonetheless, in both cases, the scaling law still holds true and hence our method can still effectively predict the effectiveness of these features.

\subsection{Limitation - when do scaling laws fail?}\label{subsec:failure-cases}

\begin{figure}[h]
  \centering
 \includegraphics[width=1.0\linewidth]{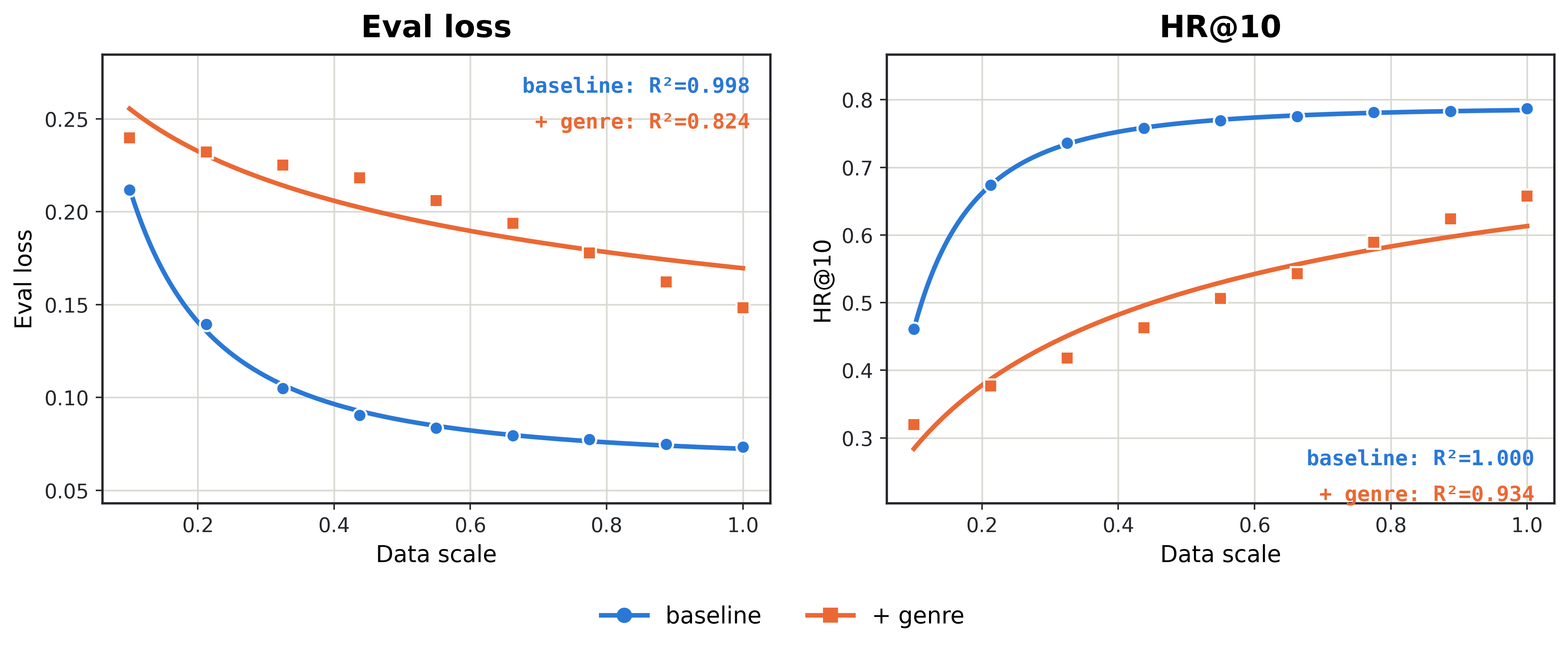}
  \caption{When genre information is included in the training of \texttt{Qwen-0.5B}, the scaling law (\textcolor{orange}{orange curve}) no longer follows a power-law fit.}
  \Description{xxx}
  \label{fig:scaling-law-failure-case}
\end{figure}

We occasionally observe that the scaling law in some settings does not follow a power-law fit. One extreme example is shown in Fig.~\ref{fig:scaling-law-failure-case} below, where the scaling law of \texttt{Qwen-0.5B} does not follow a power-law fit after we included movie genre information into the training data (\textcolor{orange}{orange curve}). In addition, this also \textit{regresses} the LLM performance significantly as compared to the baseline. We hypothesize that this happens because adding movie genre information increases the context size by around 30\%. As a result, a small LLM does not have sufficient capacity to capture the added information within context, causing the scaling behavior to break down. In general, we concede that such unusual settings would hinder the application of our method (since we are performing predictions by assuming a power-law function for the scaling curves). However, we do not expect them to appear often.




\section{Conclusion}
In this paper, we presented a method that exploits the feature-invariant scaling behavior of recommendation models to more efficiently determine the effectiveness of new feature changes. We show that different recommendation model families enjoy feature-invariant scaling laws. Then, we show that our method accurately predicts (with confidence intervals) the effectiveness of new feature changes on recommendation models, revealing numerous feature changes that improve HSTU's SOTA performance on MovieLens-20M. Our method can be used as an efficient way for practitioners to evaluate large volumes of new recommendation algorithms and features at industry-scale.


\bibliographystyle{ACM-Reference-Format}
\bibliography{software}


\newpage
\appendix

\section{Extension to nonlinear scaling laws}
\label{app:nonlinear}

Power laws describe many scaling relationships well, but other functional forms are sometimes a better fit~\cite{muennighoff2025scalingdataconstrainedlanguagemodels}. Our approach does not depend on the power-law assumption. It extends to any parametric scaling law $L_{\mathrm{feature}}(s)=f(s;\boldsymbol{\theta})$. As an example, consider the \textit{shifted power law}
\begin{equation}
    f(s;\boldsymbol{\theta})
    =
    \theta_3+\theta_1s^{-\theta_2},
    \label{eq:shifted-power-law}
\end{equation}
where $\theta_3$ is an irreducible loss floor. Because of this offset, taking logarithms no longer gives a linear model, and the posteriors lose their closed forms. We approximate them by sampling instead.

\paragraph{Model.}
As in Eq.~\eqref{eq:observation-model}, we model the observed log loss as
\begin{equation}
    y_i
    =
    \log f(s_i;\boldsymbol{\theta})
    +
    \epsilon_i,
    \qquad
    \epsilon_i
    \overset{\mathrm{i.i.d.}}{\sim}
    \mathcal{N}(0,\sigma^2).
    \label{eq:nonlinear-observation-model}
\end{equation}
Since $\theta_1$, $\theta_2$, $\theta_3$, and $\sigma$ are all positive, we run inference on their logarithms and place weakly informative Gaussian priors on them. This is equivalent to log-normal priors on the original parameters. We choose the priors to cover plausible ranges of the loss floor, the scaling exponent, and the training noise. The resulting posterior is
\begin{equation}
    p(\boldsymbol{\theta},\sigma\mid\mathcal{D})
    \propto
    p(\mathcal{D}\mid\boldsymbol{\theta},\sigma)\,
    p(\boldsymbol{\theta},\sigma),
    \label{eq:nonlinear-parameter-posterior}
\end{equation}
where the likelihood is the Gaussian implied by Eq.~\eqref{eq:nonlinear-observation-model}.

\paragraph{Posterior sampling.}
We draw $M$ samples $\{(\boldsymbol{\theta}^{(m)},\sigma^{(m)})\}_{m=1}^{M}$ from Eq.~\eqref{eq:nonlinear-parameter-posterior} using the No-U-Turn Sampler, an adaptive Hamiltonian Monte Carlo method that copes well with the strong correlations among the shifted-power-law parameters. We run several chains and check convergence using $\widehat{R}$, effective sample size, and the number of divergent transitions.

\paragraph{Full-scale prediction and decision.}
We evaluate the scaling law at the full scale $S$ for each draw:
\begin{equation}
    z_S^{(m)}
    =
    \log f\!\left(S;\boldsymbol{\theta}^{(m)}\right),
    \qquad
    m=1,\ldots,M.
    \label{eq:nonlinear-pushforward}
\end{equation}
These are samples from $p(\log L_{\mathrm{feature}}(S)\mid\mathcal{D})$, so any posterior summary can be computed from them directly. To predict a single noisy full-scale run rather than the expected loss, add $\mathcal{N}(0,(\sigma^{(m)})^2)$ noise to each draw. The posterior probability of beating the baseline is estimated as
\begin{equation}
    \widehat{p}_{\mathrm{improve}}
    =
    \frac{1}{M}
    \sum_{m=1}^{M}
    \mathbf{1}\!\left[
        z_S^{(m)}
        <
        \log L_{\mathrm{baseline}}
    \right].
    \label{eq:mc-posterior-probability}
\end{equation}
To require a minimum practical improvement $\delta$, replace the threshold with $\log\!\left((1-\delta)L_{\mathrm{baseline}}\right)$.

\section{Experiment Details: All Features}
\label{app:all-feature-description}

This section summarizes the feature and configuration changes evaluated for
HSTU (Table~\ref{tab:all-hstu-features}), Wukong, and the LLM-based recommendation model (Table~\ref{tab:all-wukong-llm-features}).

\begin{table*}[t]
\centering
\footnotesize
\setlength{\tabcolsep}{6pt}
\renewcommand{\arraystretch}{0.95}
\caption{Summary of HSTU feature and configuration changes evaluated against the baseline model.}
\label{tab:all-hstu-features}
\begin{tabular}{ll}
\toprule
\textbf{Proposal} & \textbf{Description} \\
\midrule
\texttt{rating\_additive}     & Adds a learned embedding of the observed rating to each historical item. \\
\texttt{calendar\_time}       & Adds learned hour-of-day, day-of-week, and calendar-year embeddings. \\
\texttt{item\_age}            & Adds an embedding of the movie's age at the time of the interaction. \\
\texttt{genome\_content}      & Projects the MovieLens Tag Genome vector into the item representation. \\
\texttt{embed\_dim\_15x}      & Increases the embedding and residual-stream width from 256 to 384. \\
\texttt{extra\_block}         & Adds one HSTU block, increasing the full-scale depth from 16 to 17. \\
\texttt{wider\_head\_42}      & Increases the per-head Q/K/V dimensions from 32 to 42. \\
\texttt{wider\_head\_48}      & Increases the per-head Q/K/V dimensions from 32 to 48. \\
\texttt{softmax}              & Replaces pointwise SiLU attention with scaled softmax attention. \\
\texttt{qk\_norm}             & RMS-normalizes queries and keys and applies a learned per-head logit scale. \\
\texttt{head\_temp}           & Applies a learned multiplicative scale to each head's QK logits. \\
\texttt{rope}                 & Applies rotary position embeddings to queries and keys. \\
\texttt{alibi}                & Adds a fixed head-dependent linear penalty based on relative distance. \\
\texttt{sliding\_window}      & Restricts every attention head to the 50 most recent causal positions. \\
\texttt{multi\_scale\_window} & Assigns heads cyclically to local-50, local-200, and full causal windows. \\
\texttt{attn\_res}            & Uses learned attention over the input and preceding block outputs. \\
\texttt{no\_temporal\_rab}    & Removes the temporal component of the relative attention bias. \\
\texttt{no\_positional\_rab}  & Removes the relative-position component of the attention bias. \\
\texttt{time\_gap\_bias}      & Applies a learned post-SiLU decay based on bucketized elapsed time. \\
\texttt{genre\_decay}         & Applies a post-SiLU time decay conditioned on the key item's genre. \\
\texttt{head\_time\_decay}    & Adds a learned per-head elapsed-time penalty to attention logits. \\
\texttt{rmsnorm}              & Replaces the block LayerNorm operations with learned RMSNorm. \\
\texttt{swiglu}               & Applies an additional SiLU transformation to the existing U gate. \\
\texttt{linear\_dropout\_01}  & Reduces post-gating branch dropout from 0.20 to 0.10. \\
\texttt{linear\_dropout\_025} & Increases post-gating branch dropout from 0.20 to 0.25. \\
\texttt{emb\_dropout}         & Applies additional dropout directly to item embeddings. \\
\texttt{pop\_debias}          & Subtracts a learned centered-popularity correction from candidate scores. \\
\texttt{lr\_0p1x}             & Reduces the AdamW learning rate from $10^{-3}$ to $10^{-4}$. \\
\texttt{weight\_decay\_01}    & Changes AdamW weight decay from zero to 0.1. \\
\texttt{seqlen300}            & Increases the maximum history length from 200 to 300 interactions. \\
\bottomrule
\end{tabular}
\end{table*}

\begin{table*}[t]
\centering
\footnotesize
\setlength{\tabcolsep}{5pt}
\renewcommand{\arraystretch}{0.95}
\caption{Wukong (left) and LLM-based recommendation model, Llama-3.1-8B (right) feature and configuration changes evaluated against the baseline model.}
\label{tab:all-wukong-llm-features}

\begin{minipage}[t]{0.49\textwidth}
  \centering
  \textbf{Wukong}\\[2pt]
  \begin{tabular}{@{}p{2.6cm} p{5.0cm}@{}}
    \toprule
    \textbf{Proposal} & \textbf{Description} \\
    \midrule
    \texttt{sgd\_momentum}           & Replaces the optimizer with SGD with momentum. \\
    \texttt{dense\_lr\_1e-3}         & Sets the dense-parameter learning rate to $10^{-3}$. \\
    \texttt{shampoo}                 & Replaces the optimizer with the Shampoo second-order optimizer. \\
    \texttt{drop\_one\_feature}      & Removes a single sparse input feature. \\
    \texttt{drop\_multiple\_features}& Removes several sparse input features. \\
    \texttt{dropout\_0p1}            & Adds dropout with probability 0.1. \\
    \texttt{attn\_replaces\_fmb}     & Replaces the Factorization-Machine Block with an attention module. \\
    \bottomrule
  \end{tabular}
\end{minipage}
\hfill
\begin{minipage}[t]{0.49\textwidth}
  \centering
  \textbf{LLM (Llama-3.1-8B)}\\[2pt]
  \begin{tabular}{@{}p{2.2cm} p{5.4cm}@{}}
    \toprule
    \textbf{Proposal} & \textbf{Description} \\
    \midrule
    \texttt{genre}      & Injects movie genre information into the item representation. \\
    \texttt{norating}   & Removes the observed user rating signal from the input. \\
    \texttt{dropout}    & Adds dropout regularization to the model. \\
    \texttt{extralayer} & Adds one additional transformer decoder layer, increasing depth. \\
    \bottomrule
  \end{tabular}
\end{minipage}
\end{table*}

\section{Full Results for All Features}
\label{app:all-result}

\begin{figure*}[h] \centering \includegraphics[width=0.9\textwidth]{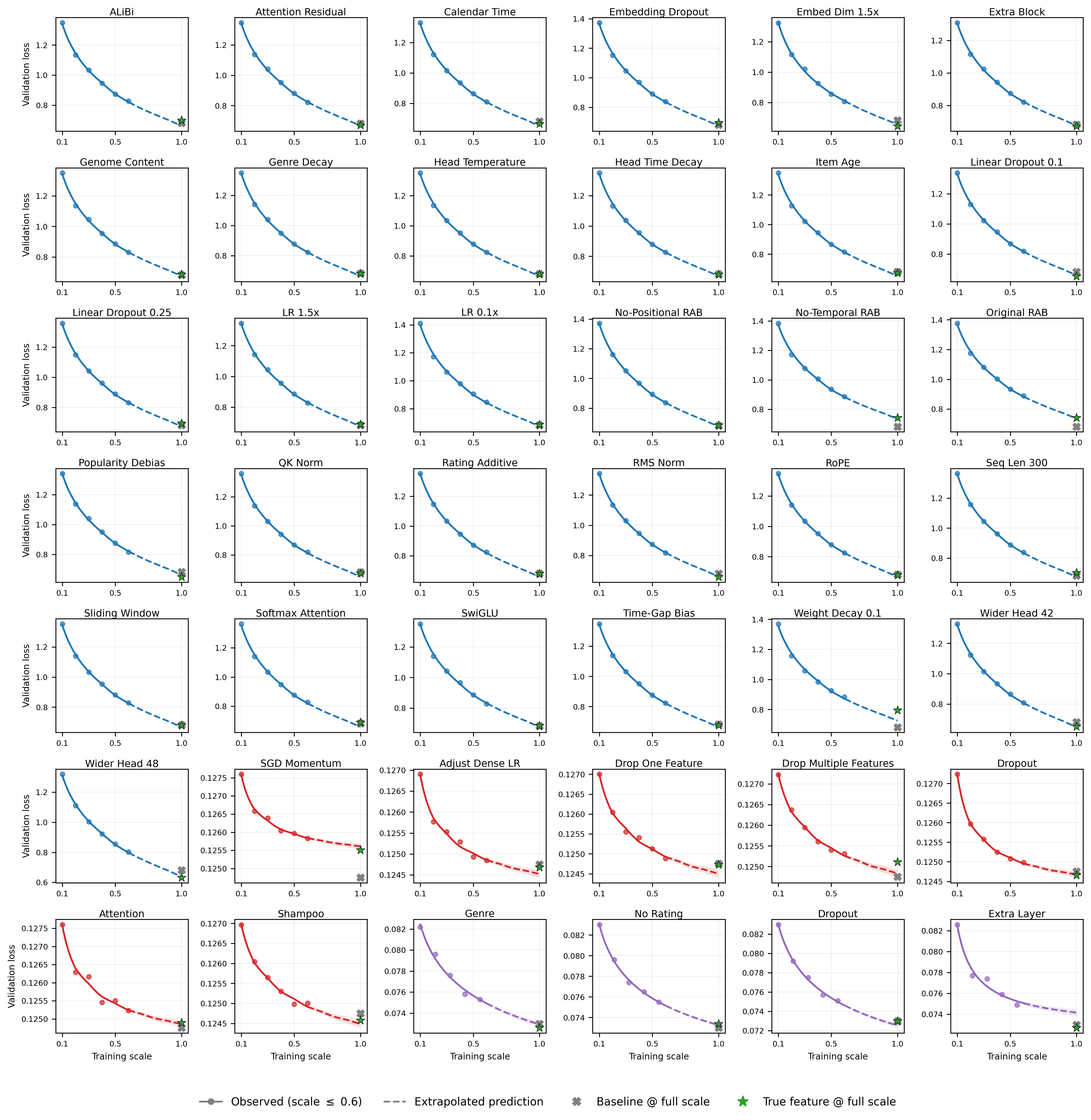}
\caption{Prediction of full-scale experimental model performance for all features and models. \textcolor{blue}{Blue}: HSTU. \textcolor{red}{Red}: Wukong. \textcolor{purple}{Purple}: Llama-3.1-8B-Instruct.}
\label{fig:extrapolation_all_features}
\end{figure*}

Fig.~\ref{fig:extrapolation_all_features} shows the full-scale predictions for every feature, and Table~\ref{tab:all_probability_sorted} reports the posterior probability of improvement as more small-scale observations are added.

\begin{table*}[t]
\centering
\caption{Probability that a feature's full-scale validation loss is lower than the baseline model by a percentage margin
(HSTU at the 1\% margin; Wukong and LLM at 0.05\% since their effects are sub-1\%).
Rows ordered by true full-scale eval-loss change, best to worst.}
\label{tab:all_probability_sorted}
\footnotesize
\setlength{\tabcolsep}{3pt}
\renewcommand{\arraystretch}{0.95}

\begin{minipage}[t]{0.5\textwidth}
\centering
\textbf{HSTU}\\[2pt]
\begin{tabular}{lrrrr}
\toprule
\textbf{Feature} & \textbf{0.1--0.4} & \textbf{0.1--0.5} & \textbf{0.1--0.6} & \textbf{Full} \\
\midrule
Wider Head 48       & 98.7\% & 98.7\% & 99.1\% & $-7.3\%$ \\
Embed Dim 15X       & 91.2\% & 91.2\% & 93.2\% & $-5.4\%$ \\
Pop Debias          & 90.2\% & 90.5\% & 90.9\% & $-4.8\%$ \\
Linear Dropout 01   & 90.7\% & 91.3\% & 93.4\% & $-4.5\%$ \\
Wider Head 42       & 96.1\% & 96.0\% & 97.9\% & $-4.5\%$ \\
Rmsnorm             & 90.6\% & 91.6\% & 96.0\% & $-3.4\%$ \\
Calendar Time       & 92.1\% & 92.2\% & 94.9\% & $-2.1\%$ \\
Attn Res            & 64.6\% & 90.8\% & 93.3\% & $-1.5\%$ \\
Extra Block         & 67.1\% & 72.9\% & 78.1\% & $-1.3\%$ \\
Item Age            & 90.7\% & 91.6\% & 92.7\% & $-1.2\%$ \\
Qk Norm             & 95.3\% & 96.0\% & 95.4\% & $-1.1\%$ \\
Time Gap Bias       & 78.8\% & 79.9\% & 81.5\% & $-0.8\%$ \\
Rope                & 75.4\% & 73.9\% & 75.0\% & $-0.6\%$ \\
Rating Additive     & 90.2\% & 93.5\% & 91.7\% & $-0.4\%$ \\
Sliding Window      & 79.0\% & 73.9\% & 70.5\% & $-0.3\%$ \\
Genre Decay         & 76.0\% & 78.2\% & 82.6\% & $-0.2\%$ \\
Head Temp           & 73.3\% & 70.5\% & 71.3\% & $-0.2\%$ \\
Head Time Decay     & 62.8\% & 62.8\% & 63.3\% & $-0.2\%$ \\
Swiglu              & 40.6\% & 37.4\% & 45.7\% & $+0.2\%$ \\
Genome Content      & 57.6\% & 47.6\% & 49.2\% & $+0.6\%$ \\
Lr 0P1X             & 76.0\% & 59.9\% & 51.8\% & $+0.6\%$ \\
No Positional Rab   & 53.0\% & 44.5\% & 44.8\% & $+0.6\%$ \\
Lr1.5X              & 49.2\% & 44.0\% & 51.9\% & $+0.8\%$ \\
Softmax             & 92.2\% & 91.2\% & 87.1\% & $+1.0\%$ \\
Linear Dropout 025  & 33.5\% & 24.4\% & 18.0\% & $+1.6\%$ \\
Emb Dropout         & 39.9\% & 24.8\% & 17.7\% & $+2.2\%$ \\
Seqlen300           & 36.1\% & 31.4\% & 12.5\% & $+3.0\%$ \\
Original Rab        &  2.4\% &  0.5\% &  0.1\% & $+8.6\%$ \\
No Temporal Rab     &  4.6\% &  1.1\% &  0.3\% & $+8.7\%$ \\
Weight Decay 01     & 17.7\% &  3.9\% &  1.0\% & $+16.6\%$ \\
\bottomrule
\end{tabular}
\end{minipage}
\hfill
\begin{minipage}[t]{0.48\textwidth}
\centering
\textbf{Wukong}\\[2pt]
\begin{tabular}{lrrrr}
\toprule
\textbf{Feature} & \textbf{0.1--0.4} & \textbf{0.1--0.5} & \textbf{0.1--0.6} & \textbf{Full} \\
\midrule
Shampoo                & 100.0\% & 100.0\% & 100.0\% & $-0.12\%$ \\
Dropout                & 91.4\%  & 99.9\%  & 98.9\%  & $-0.10\%$ \\
Adjust Dense LR        & 0.1\%   & 99.3\%  & 100.0\% & $-0.10\%$ \\
Drop One Feature       & 2.9\%   & 83.8\%  & 100.0\% & $-0.08\%$ \\
Attention              & 92.2\%  & 0.0\%   & 0.0\%   & $+0.1\%$ \\
Drop Multiple Features & 86.6\%  & 90.7\%  & 0.0\%   & $+0.3\%$ \\
SGD Momentum           & 0.0\%   & 0.0\%   & 0.0\%   & $+0.6\%$ \\
\bottomrule
\end{tabular}

\vspace{4mm}

\textbf{LLM (Llama-3.1-8B)}\\[2pt]
\begin{tabular}{lrrrr}
\toprule
\textbf{Feature} & \textbf{0.1--0.4} & \textbf{0.1--0.5} & \textbf{0.1--0.6} & \textbf{Full} \\
\midrule
+ Genre     & 12.7\% & 60.2\%  & 91.1\%  & $-0.5\%$ \\
+ Rating    & 43.0\% & 100.0\% & 100.0\% & $-0.5\%$ \\
Extra Layer & 71.5\%  & 90.0\%   & 90.9\%   & $-0.4\%$ \\
Dropout     & 0.4\%  & 0.0\%  & 9.8\%  & $+0.0\%$ \\
\bottomrule
\end{tabular}
\end{minipage}
\end{table*}

\end{document}